# A NOVEL APPROACH FOR AUTHENTICATING TEXTUAL OR GRAPHICAL PASSWORDS USING HOPFIELD NEURAL NETWORK

ASN Chakravarthy[†], Prof. *P S Avadhani*[††], P. E. S. N Krishna Prasad[†††], N.Rajeev[††††] and D.Rajasekhar reddy[†††††]

[†] Associate Professor , Department of CSE& IT, Sri Sai Aditya Institute of Science and Technology, Surampalem, E.G.Dist , Andhra Pradesh, India
`asnchakravarthy@yahoo.com`

[††]Professor, Dept. of CS & SE, Andhra University, Visakhapatnam Dist, Andhra Pradesh, India
`psavadhani@yahoo.com`

[†††]Department of CSE, Aditya Engineering College, Kakinada, India.
`surya125@gmail.com`

[††††]Final year B.Tech , Department of CSE, Sri Sai Aditya Institute of Science and Technology, Surampalem, E.G.Dist , Andhra Pradesh, India
`nakka.rajeev@gmail.com`

[†††††]Final year B.Tech , Department of CSE, Sri Sai Aditya Institute of Science and Technology, Surampalem, E.G.Dist , Andhra Pradesh, India
`raj234057@gmail.com`

## ABSTRACT

*Password authentication using Hopfield Networks is presented in this paper .In this paper we discussed the Hopfield Network Scheme for Textual and graphical passwords, for which input Password will be converted in to probabilistic values. We observed how to get password authentication using Probabilistic values for Textual passwords and Graphical passwords. This study proposes the use of a Hopfield neural network technique for password authentication. In comparison to existing layered neural network techniques, the proposed method provides better accuracy and quicker response time to registration and password changes.*

## 1. INTRODUCTION

At the advent of the information explosion and the increasing users of Internet, the data transmission has become vital in every activity of human life. This necessitated the need to protect data during the transmission and checking the data for its authenticity and also the authenticity of the sender and receiver. In this context, authentication of the user has become a prime concern among many access control mechanisms. Secure networks allows only intended recipient to intercept and read a message addressed to him. Thus protection of information is required against possible violations than compromise its secrecy. Secrecy is compromised if information is disclosed to users not authorized to access it. Password authentication is one of the mechanisms that are widely used to authenticate an authorized user . Authentication is the act of confirming the truth of an attribute of a datum or entity. This might involve confirming the identity of a person, tracing the origins of an artefact, ensuring that a product is what it's packaging and labelling claims to be, or assuring that a computer program is a trusted one. Computer system that is supposed to be used only by those authorized must attempt to detect and exclude the unauthorized. Access to it is therefore usually controlled by insisting on an authentication procedure to establish with some degree of confidence the identity of the user, thence granting those privileges as may be authorized to that identity.

## 1.1 RELATED WORK

In a paper on Password Authentication Using Hopfield Neural Networks by Shouhong Wang and Hai Wang [1] proposed a new password authentication scheme based on HNN. Without losing its generalities, they are able to demonstrate this authentication scheme to verify user IDs and passwords. This authentication scheme can be used for any access control of computing resources, such as multiple server access permission [2] or role-based security control [3]. This authentication scheme includes three major procedures: registration, log-in authorization, and password change. This method purely based on taking textual passwords as input. Since Textual passwords suffer from replay attack the level security degrades. In our proposed scheme the textual password is normalized and we are taking probabilistic values as input to HPNN method. In order to have more authentications this scheme is extended for graphical passwords.

## 2. HOPFIELD NEURAL NETWORKS (HPNN):

### 2.1 Hopfield model:

The Hopfield model[4,5] is a distributed model of an associative memory. The Hopfield Model was proposed by John Hopfield of the California Institute of Technology during the early 1980s. The dynamics of the Hopfield model is different from that of the Linear Associator Model in that it computes its output recursively in time until the system becomes stable. We presented a Hopfield model with six units, where each node is connected to every other node in the network is given in figure 1. Unlike the linear associator model which consists of two layers of processing units, one serving as the input layer while the other as the output layer, the Hopfield model consists of a single layer of processing elements where each unit is connected to every other unit in the network other than itself. The connection weight matrix W of this type of network is square and symmetric, i.e., $w_{ij} = w_{ji}$ for i, j = 1, 2... m. Each unit has an extra external input $I_i$. This extra input leads to a modification in the computation of the net input to the units.

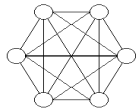

*Figure1. Hopfield Network*

Unlike the linear associator, the units in the Hopfield model act as both input and output units. But just like the linear associator, a single associated pattern pair is stored by computing the weight matrix as follows:

$$W_k = X_k^T Y_k, \quad where\ Y_k = X_k$$

(1)

$$W = \alpha \sum_{k=1}^{p} W_k$$

to store p different associated pattern pairs. Since the Hopfield model is an auto-associative memory model, patterns, rather than associated pattern pairs, are stored in memory. After encoding, the network can be used for decoding. Decoding in the Hopfield model is achieved by a collective and recursive relaxation search for a stored pattern given an initial stimulus pattern. Given an input pattern X, decoding is accomplished by computing the net input to the units and determining the output of those units using the output function to produce the pattern X'. The pattern X' is then fed back to the units as an input pattern to produce the pattern X". The pattern X" is again fed back to the units to produce the pattern X'''. The process is repeated until the network stabilizes on a stored pattern where further computations do not change the output of the units. During decoding, there are several schemes that can be used to update the output of the units. The updating schemes are Synchronous (parallel), Asynchronous (sequential), or a combination of the two (hybrid). Using the synchronous updating scheme, the output of the units is updated as a group prior to feeding the output back to the network. On the other hand, using the asynchronous updating scheme, the output of the units are updated in some order (e.g. random or sequential) and the output are then fed back to the network after each unit update. Using the hybrid synchronous-asynchronous updating scheme, subgroups of units are updated synchronously while units in each subgroup updated asynchronously. The choice of the updating scheme has an effect on the convergence of the network. Hopfield (1982) demonstrated that the maximum number of patterns that can be stored in the Hopfield model of m nodes before the error in the retrieved pattern becomes severe is around 0.15m. The memory capacity of the Hopfield model can be increased as shown by André cut (1972). Hopfield model is broadly classified into two categories: Discrete Hopfield Model and Continuous Hopfield Model.

## 3. AUTHENTICATION USING HOPFIELD NEURAL NETWORK:

Working of this technique is similar to feed forward network but here rapid training is possible. This method takes set of patterns as input and performs the training. Once the training has been completed the network will be tested. In testing it takes a pattern as input and produces the output. If input pattern and output pattern are same then it means that the pattern is recalled. So it is not possible to train network by giving input and corresponding output pairs. In order to overcome this problem we will merge input (user name) and output (password) and we will give this merged pattern as input to the network.

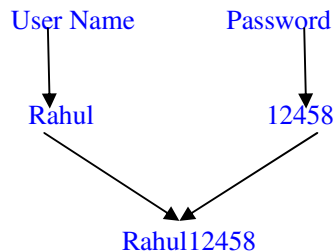

*Figure2. Merging the Username and Password*

But merging of username and password leads to some other problems such as merging of two different input output pairs may lead to single pattern.

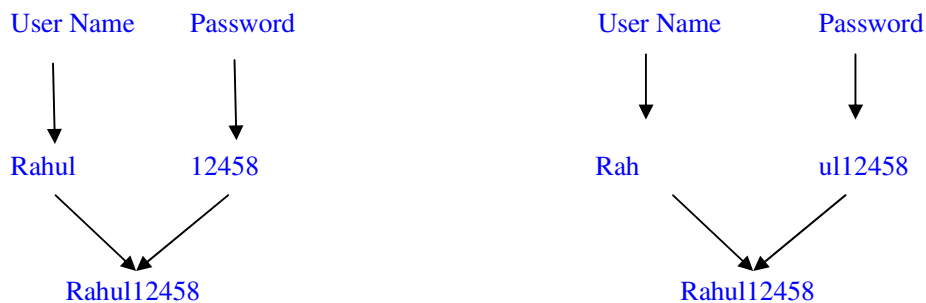

*Figure3. Problem with merging*

To overcome above problem we have to use an additional character as delimiter between username and password.

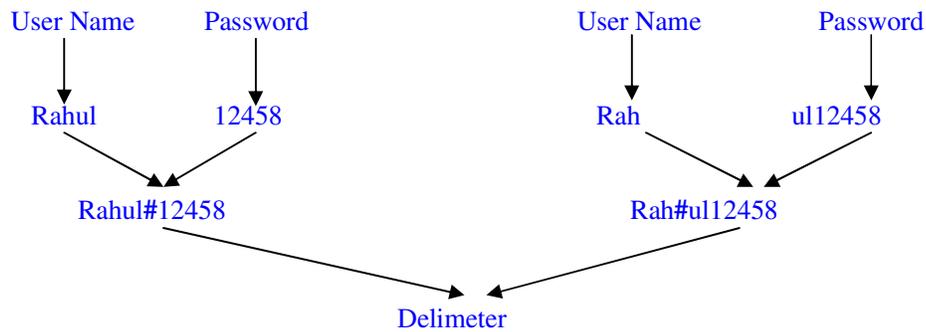

*Figure4. Adding Delimiter for Merging Process*

## 3.1 Authentication Procedure Using HP:

The password authentication using Hopfield method may use any one of the Textual or Graphical password as password and can train network so that it can authenticate users.

### 3.1.1 Textual password

Initially this method will convert username and password into binary values and we can use those values as training samples.

This method performs following steps.

- Convert each character into a unique number (for example ASCII value)
- Convert the unique number into binary value

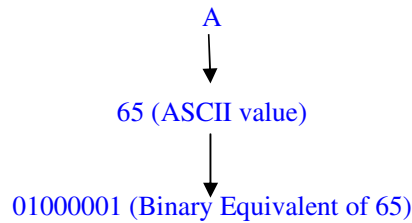

*Figure5. Convert the unique number into binary value*

By using above procedure all characters in the username and password will be converted into binary values.

*Table1. Representing User Name in Binary Values*

| Username | Binary value representing username |
|---|---|
| MANIDEEP | 01011001010000100111001010010010001000101010001010100010000101 |
| HARSHA | 000010010100000100100101011001010000100101000001 |
| SATISH | 011100010101000100010101010010010011001 |

After converting username and password into binary equivalents we will use these pairs as training samples. Once the training has been completed we will store network in each server. When user wants to get service from a server he/she submits user name and password to the server, then server loads network and generates output by giving username as input. If the output matches with the password submitted by the user then server provides service. This scheme provides better authentication by using bipolar input instead of binary input.

This method converts a binary number into bipolar number by using following formula or by simply replacing zeros with -1s. If X is a binary digit then corresponding bipolar value is (2X-1).

$$1 \rightarrow 1$$
$$0 \rightarrow -1$$

By using above procedure we can convert binary value in to bipolar value and we can use it as input to the network.

*Table2. Converting Binary Value to Bipolar Value*

| Username | Binary value representing username |
|----------|-------------------------------------|
| MANIDEEP | -11-111-1-11-11-1-1-1-1-11-1-1111-1-11-11-1-11-1-11-1-1-11-1-1-11-11-11-1-1-11-11-11-1-1-11-1-1-1-1-11-11 |
| HARSHA   | -1-1-1-11-1-11-11-1-1-1-1-11-1-11-1-11-11-111-1-11-11-1-1-1-11-1-11-11-1-1-1-1-11 |
| SATISH   | -111-1-11-11-11-1-1-1-1-11-1-1-11-11-11-1-11-1-11-111-1-11-11-1-1-1-11-1-11 |

**3.1.2 Graphical password**

Giving image directly as input to the neural network is not possible. So here we have to convert image into text. Proposed scheme can do this in the following way

**3.1.3 Image to RGB Conversion**

This scheme reads colour of each pixel of the selected image and converts the colour into red, green and blue (RGB) parts as any colour can be produced using these three primary colours as shown in figure 6.

**3.1.4 Converting Image to Binary Values:**

The above scheme first converts the image into matrix, representing RGB values. Then it converts the RGB values into binary values and represents in the form of a matrix. As each value (R, G and B) ranges from 0 to 255 we can assign value using a 8-bit binary number.

$$135 \quad 206 \quad 235$$
$$\downarrow \quad \downarrow \quad \downarrow$$
$$10000111 \quad 11001110 \quad 11101011$$

*Figure7. Converting RGB to Binary Values*

The above procedure produces the matrix consisting of RGB values into matrix consisting of binary values representing all the pixels of the image.

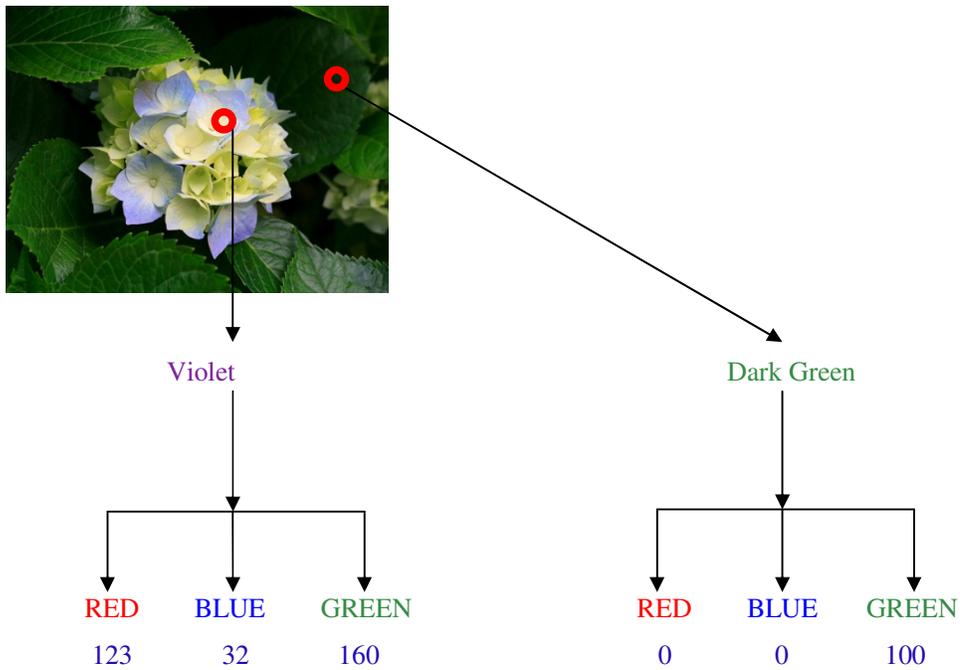

Figure6. Converting Image to RGB Value

By practicing the above custom we can convert any image into a matrix enduring of set of numbers representing all the pixels of the image.

$$\begin{bmatrix} 1 & 0 & 0 & 0 & 0 & 1 & 1 & 1 & 1 & 1 & 0 & 0 & 1 & 1 & 1 & 0 & 1 & 1 & 1 & 0 & 1 & 0 & 1 & 1 \\ \dots & & & & & & & & & & & & & & & & & & & & & & & \\ 1 & 0 & 0 & 1 & 0 & 1 & 0 & 0 & \dots & & & & & & & & & & & & & & & \end{bmatrix}$$

Figure8. Matrix Representing Pixels of the Image

### 3.1.5 Converting Image to Bipolar Values

The above procedure first converts the image into a matrix representing binary values and then converts the binary values into bipolar values by replacing 0 with -1 and represent in the form of a matrix.

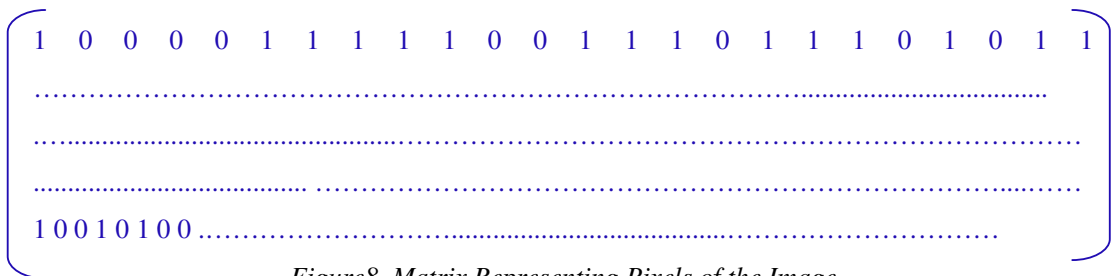

By utilizing the above layout this scheme converts the matrix consisting of binary values into matrix consisting of bipolar values representing all the pixels of the image as shown in figure 9.

After converting image into bipolar values we can use the same procedure which is used for textual password.

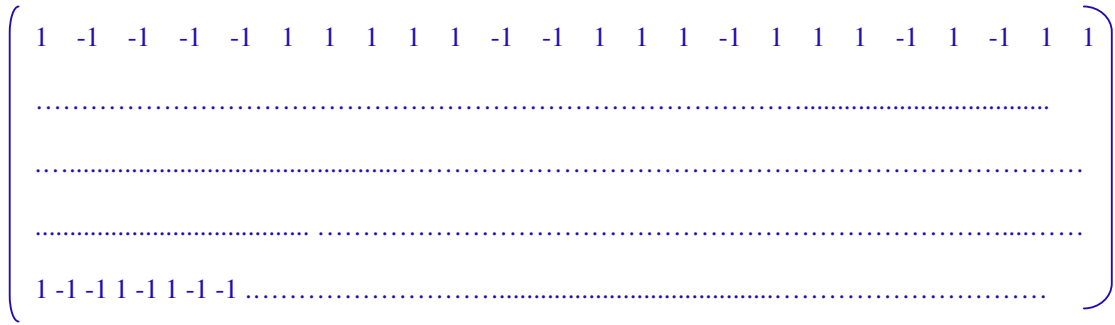

*Figure9. Matrix representing bipolar values of all the pixels*

### 3.2 Learning:

When a new user creates an account, network has to adjust weights so that it can recognize all the users who are registered. This process of changing weights is called learning.

#### 3.2.1 Learning in Hopfield Neural Network:

Hopfield network takes a pattern as input and adjusts its weights. When we give same pattern as input it already which is already stored, it will be recalled. Suppose $(u_1, p_1)$ $(u_2, p_2)$ ,…………………..,$(u_n, p_n)$ are username password pairs we want to store in the network. First this method will merge these pairs as discussed in above sections and converts these merged patterns into binary or bipolar format. Let $R_1, R_2, ……………., R_n$ be the pattern after merging. Each binary digit can be converted into bipolar value by using the following formula or by simply replacing zeros with -1s. Let X is a binary digit, and then its corresponding bipolar value is (2X-1).

$$0 \rightarrow -1$$
$$1 \rightarrow 1$$

Let X1, X2,……………., Xn be the bipolar patterns representing $R_1, R_2, ………, R_n$, the weight matrix M can be calculated by using the equation 2.

$$M = \sum_{i=0}^{n} X_i^T X_i \qquad (2)$$

### 4. IMPLEMENTATION DETAILS:

int NoOfPatterns = 0;
int NoOfBitsPerPattern = 0;
int[,] Weight,Input;

Here in this scheme NoOfPatterns: Indicates the no of patterns we want to use in training, NoOfBitsPerPattern: Specifies no of bits we want to use for each pattern, Weight: Stores weight values of the network and Input: Stores input vector.

### 4.1 Implementing HPNN Training:

Before training, application will take training samples from the user and stores them in the corresponding variables.

```
    private void Train()
    {
```

Advanced Computing: An International Journal ( ACIJ ), Vol.2, No.4, July 2011

```
    for (int i = 0; i < Input.GetLength(0); i++)
    {
        int[,] pattern = MatrixMath.GetRow(Input, i);
      int[,] temp = MatrixMath.Multiply(MatrixMath.Transpose(pattern), pattern);
        int[,] IdMatrix=MatrixMath.GetIdentityMatrix(temp.GetLength(0));
        int[,] ContributionMatrix=MatrixMath.Subtract(temp,IdMatrix);
        Weight=MatrixMath.Add(Weight, ContributionMatrix);
    }
}
```

**Note:** Here instead of calculating the transpose we multiplied the required elements which will be multiplied when we calculate transpose.

### 4.2 Recognizing the Pattern using HPNN:

The pattern which we want to use for testing the network will be endowed as input to the application and then application stores the pattern in the corresponding variable.

**This is implemented as follows**

```
private void toolStripButton4_Click(object sender, EventArgs e)
{
  try
  {
    int[,] pattern = CreateMatrix(tableLayoutPanel2);
    int[,] result = MatrixMath.Multiply(Weight, pattern);
    result = MatrixMath.GetBipolar(result);
    MessageBox.Show(MatrixMath.GetString(result));
  }
  catch (Exception ee) { MessageBox.Show(ee.Message); }
}
```

## 5. RESULTS FOR HPNN:

### 5.1 HPNN for Textual Passwords:

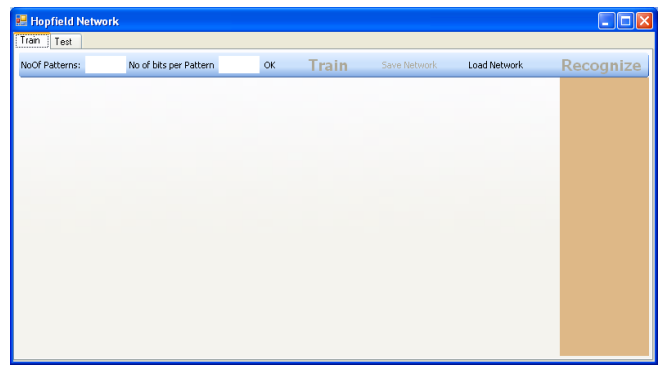

*Figure10. Screen showing how to setup network*

In this screen the No of Patterns indicates the number of patterns that we want to use in training and No of Bits per Pattern specifies the number of bits that we want to use for each pattern.

Once the required information has given OK button will be pressed then the application will provide enough fields to enter input and output pairs.

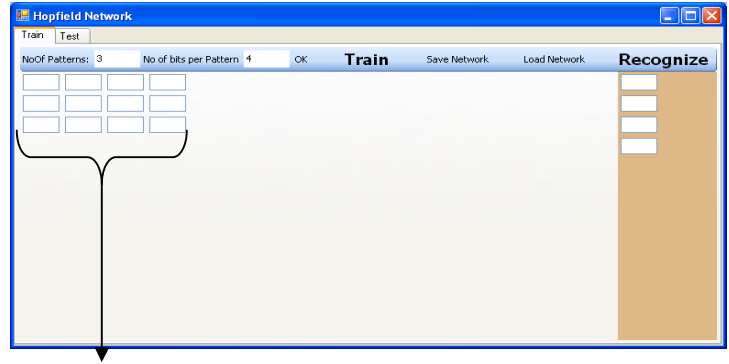

**Input fields**

*Figure11. Screen showing Number of Patterns*

### 5.1.1 Training the HPN Network:

Once the required training set has given, and the Train button is pressed the application starts training for the given input values.

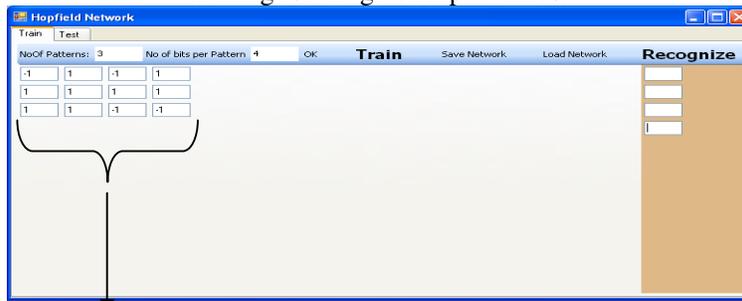

**Training Set**

*Figure12. Screen showing HPNN training*

### 5.1.2 User Authentication Using HPNN

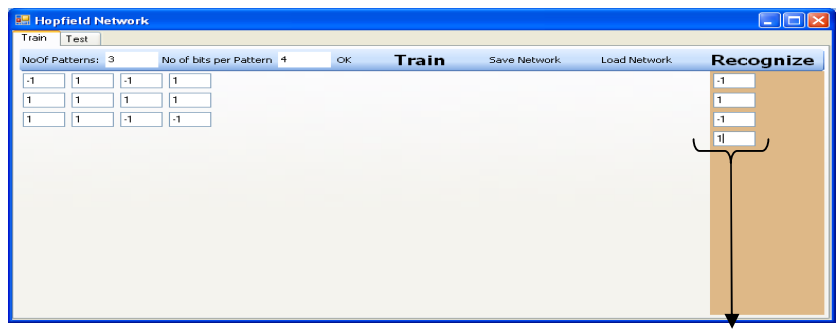

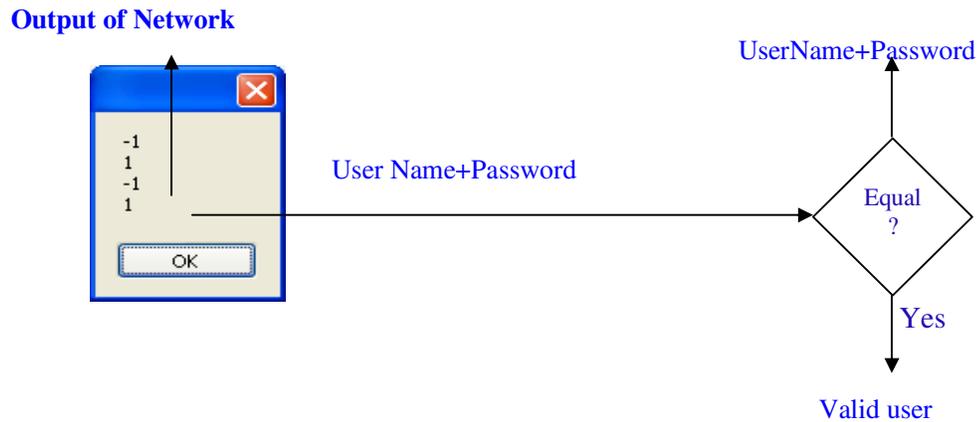

*Figure13. Screen showing User Validation Using HPNN*

The application compares the output of the network with the password given by the user as shown in figure 13, if both are the same then the user is valid and he will be connected by the server.

## 5.2 HPNN for Graphical Passwords:

Giving an image directly as input to the network is impossible. So here the application converts the selected image into text.

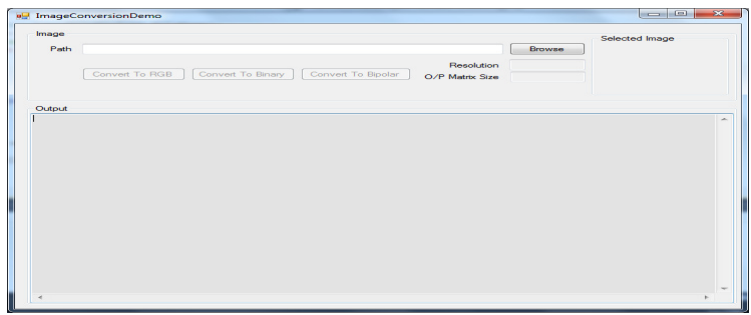

*Figure14. Screen showing how to select an the image*

In figure 14 the *Path* denotes path of the image we want to convert, *Selected Image* shows the selected image ,*Resolution* specifies resolution of the selected image and *O/P Matrix Size* stipulates size of the output matrix.

After an image has been selected it will be disclosed in the *Selected Image* box and conversion buttons will be facilitated. The aplictaion provides a faclity to select an image from any one of the folders in the system to select required image as an input . Once the image is selecetd it will be displayed as a thumb nail and resoultion is also dispalyed as shown in figure 15.

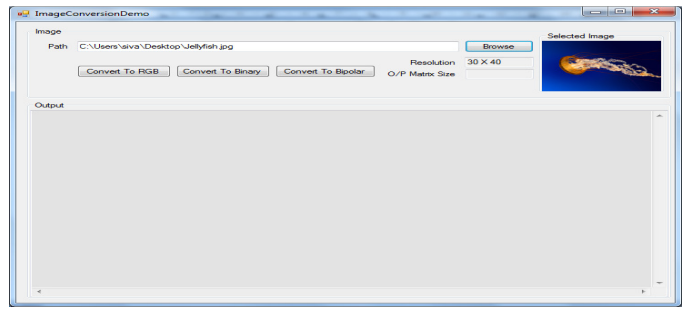

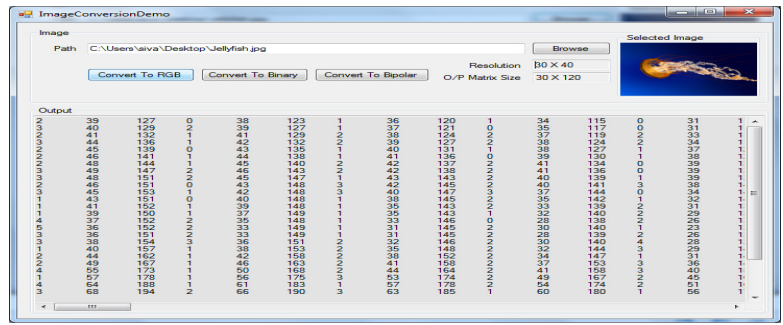

*Figure15. Screen showing how to convert an image into an Integer (RGB) matrix*

*Figure16. Output matrix*

After electing an image "Convert To RGB" is pressed then we can monitor the output matrix in output box in figure 16 and also it displays the size of the output matrix. The sceern shows the integer values of an image.

These integer values will be converted in to binary values so as to satisfying our probablistic method,where the input to the network should be normalized values .

After selecting an image if "Convert To Binary "button is pressed then we can scrutinize the output matrix in the output box as shown in figure 17.

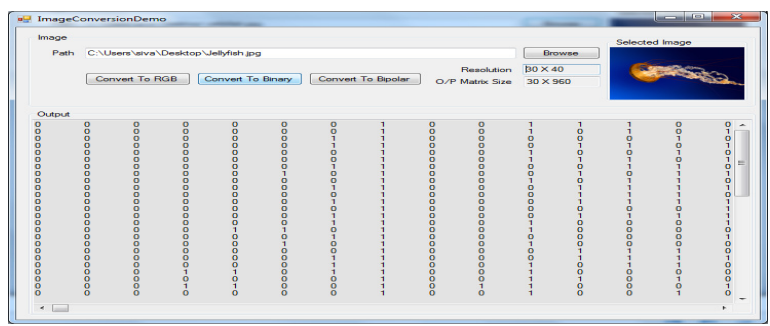

Figure 17. Screen showing how to convert an image into a binary matrix

This screen shows all the binary values corresponding to the inateger values of the image which has been selected through the application. These binary values can be used for the trainig the Hopfiled network for graphical password authentication.

The application also converts the binary values in to bipolar values after pressing the button" "Convert To Bipolar". Then we can perceive output matrix in output box.

43

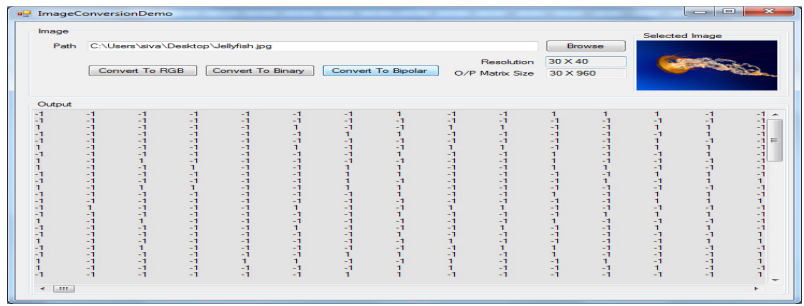

*Figure18. Screen showing how to convert an image into a bipolar matrix*

In the above classification we have seen how to convert an image into text. Once the image has been converted we can use it as normal password

## 6. CONCLUSION

Access authentication is crucial for computer security. The use of neural networks has been proposed to eliminate the defective features the traditional Verification table approach. However, the existing layered neural network method suffers several limitations such as the lengthy training time and the arbitration in authentication. This paper shows that a probabilistic approach for HNN-based authentication scheme can effectively be used for access authentication in the open computing environment.

This Paper showed that an HNN-based authentication scheme can effectively be used for access authentication in the open computing environment. An HNN, with large capacity, can store authentication information using marginal training time. The authentication scheme incorporating the use of HNN can recall information for a legal user's ID and password instantly and accurately. Our experiments have demonstrated the usefulness and robustness of the proposed authentication scheme.

Table 3. Performance Comparison

| Number of users | Hopfield network | Layered architecture |
|---|---|---|
| **25** | 0.000435 sec | 91.00 sec |
| **50** | 0.000794 sec | 317 .0sec |
| **100** | 0.00136 0sec | 1876 sec |
| **10 million** | 213.0000 sec | Computational cost is high |

Table 4.Training times of different networks for different inputs

| | Training Time | | |
|---|---|---|---|
| **BPNN** | 360 | 450 | 500 |
| **HPNN** | 136 | 50 | 100 |

Advanced Computing: An International Journal ( ACIJ ), Vol.2, No.4, July 2011

## 6. FUTURE SCOPE

Enhancement can be provided for the authentication by using alternate active functions like Hyperbolic Tangent, Linear, SoftMax, Tangential, Sin Wave, Bipolar and Gaussian etc. In the proposed work a password authentication scheme using associative memories based on normalized combined text and graphical passwords can be used. A virtual keypad can be provided through which password can be entered and can define some special characters in the character set for text passwords. For graphical passwords we can draw images or symbols on the virtual screen and can use those images as passwords.

## ACKNOWLEDGEMENTS

An assemblage of this nature could never have been attempted without reference to and inspiration from the works of others whose details are mentioned in reference section.

## Authors

†A.S.N Chakravarthy received his M.Tech (CSE) from JNTU, Anantapur , Andhra Pradesh, India. Presently he is working as an Associate Professor in Computer Science and Engineering in Sri Sai Aditya Institute of Science Technology, SuramPalem, E.G.Dist, AP, India. He is a Life member of CSI, ISCA. He has 06 papers published in various National / International journals and conferences His research area includes Network Security, Cryptography, Intrusion Detection, Neural networks.

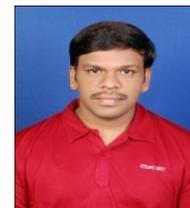

††Prof. P.S.Avadhani did his Masters Degree and PhD from IIT, Kanpur. He is presently working as Professor in Dept. of Computer Science and Systems Engineering in Andhra University college of Engg., in Visakhapatnam. He has more than 50 papers published in various National / International journals and conferences. His

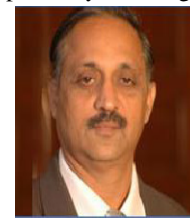

Advanced Computing: An International Journal ( ACIJ ), Vol.2, No.4, July 2011

research areas include Cryptography, Data Security, Algorithms, and Computer Graphics, Digital Forensics and Cyber Security.

††† P E S N Krishna Prasad, currently is a Research Scholar under the guidance of Dr. BDCN Prasad in the area of Machine Intelligence and Neural Networks. He is working as Associate Professor in the Department of CSE, Aditya Engineering College, Kakinada, Andhra Pradesh, India. He is a member of ISTE. He has presented and published papers in several national and International conferences and journals. His areas of interest are Artificial Intelligence, Neural Networks and Machine Intelligence.

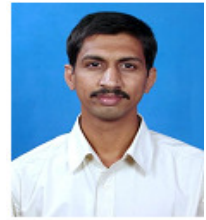

††††N Rajeev studying his B.Tech (CSE) in Sri Sai Aditya Institute of Science Technology, SuramPalem, E.G.Dist, AP, India.

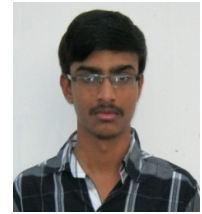

†††††D Rajasekhar reddy studying his B.Tech (CSE) in Sri Sai Aditya Institute of Science Technology, SuramPalem, E.G.Dist, AP, India.

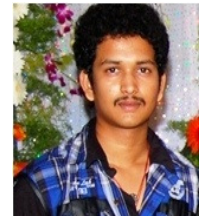

o0o